# Accessible 3D Printed Archimedes System for Accurate and Precise Density Measurements and Uncertainty Analysis

Emily Doucet[1], Jan-Hendrik Pöhls[1,2,*]

1) Department of Chemistry, University of New Brunswick, Fredericton, NB, Canada
2) Department of Physics, University of New Brunswick, Fredericton, NB, Canada

* Corresponding author: Jan.Pohls@UNB.ca

**Abstract:** Accurate and precise density measurements are important for characterizing solid materials and for evaluating physical properties that are strongly affected by porosity. While several methods are available for determining density, measurements of small or irregularly shaped samples can be limited by sample geometry, balance precision, liquid density, and other experimental sources of uncertainty. In this study, an accessible Archimedes density measurement system was developed using 3D printed components and an analytical balance for the measurement of solid and liquid densities. The measurement of solid densities requires knowing the exact density of the liquid. Therefore, copper, vanadium, and indium standards were used to measure the densities of ethanol ($0.8116 \pm 0.0005$ g/cm$^3$), distilled water ($0.9982 \pm 0.0013$ g/cm$^3$), and hexanes ($0.6748 \pm 0.0016$ g/cm$^3$), all of them showing strong agreement with literature values. Uncertainty analysis indicates that the relative uncertainty in the measured solid density decreases with increasing sample mass and balance precision, while higher density liquids can further reduce the required sample mass. The uncertainty of the immersed mass measurement was found to have a greater influence on the calculated density than the uncertainty of the air mass measurement. An open-source Archimedes LS software was developed to analyse liquid and solid densities including the corresponding uncertainties. These findings demonstrate that an accessible 3D printed Archimedes system combined with an open-source software provides a practical approach for improving the consistency and comparability of density measurements between laboratories.

# Introduction

Density, ρ, is a fundamental intrinsic physical property that provides a material's mass-to-volume relationship and can be used to evaluate the compactness of a solid sample. Dense materials often exhibit enhanced mechanical properties such as hardness, stiffness, and compressive strength compared to their porous counterparts.[1,2] There are three different definitions of density: the intrinsic – also theoretical – density can be determined from the crystal structure by dividing the atomic weight in the unit cell by its volume; the skeletal density, which accounts only for the solid material and excludes open surface pores; and the actual (apparent) density which includes both the solid material and the effects of internal and surface-accessible pores. While the actual density of single crystals is often very close to the theoretical density, it is rarely equal to the value due to the presence of vacancy and interstitial sites, dislocations, or isotopes. For example, isotopically enriched $^{12}C$ diamonds have a slightly lower density than natural diamonds.[3]

Most materials, however, are polycrystalline and can therefore exhibit actual or skeletal densities that diverge substantially from their theoretical values. Polycrystalline samples are commonly prepared from powders, in which the individual grains must be efficiently packed to minimize pores and other internal voids within the consolidated material.[4,5] Hot pressing and spark plasma sintering are widely used to increase the density of the material by improving grain rearrangement and consolidation and hence, reducing the volume fraction of pores and voids. However, accurately measuring the actual or skeletal density remains important, as residual porosity can significantly influence the physical properties of the material. For example, mechanical properties in steel can be significantly reduced by the presence of pores.[6] By having 2% finely dispersed pores, the ductility of the steel was reduced from 10% to 3%. This is also significant when characterizing thermoelectric materials, as porosity can drastically affect their electrical and thermal transport properties.[7,8] Pores can, for instance, interrupt heat transport through the material and increase phonon scattering, which can lower the measured thermal conductivity.[8,9,10] The effect of porosity on thermal conductivity is therefore important to consider when comparing thermoelectric materials or interpreting their measured properties. Similarly, porosity reduces electrical conductivity by disrupting continuous electrical conduction pathways through the material[11], an effect that can be described using effective-medium models such as the Maxwell model.[12] For example, in porous SiC ceramics, increasing the porosity from approximately 30% to 63% was reported to decrease the electrical conductivity from 7.7 to 1.7 $\Omega^{-1}$ $cm^{-1}$, while the thermal conductivity simultaneously decreased from 37.9 to 5.8 W $m^{-1}\cdot K^{-1}$.[11] While it is often assumed that the ratio of electrical and thermal conductivity remains the same, this example indicates that the ratio changes from 20.3 $K\cdot V^{-2}$ to 29.3 $K\cdot V^{-2}$. Because even small amounts of porosity can result in substantial changes in physical properties, accurate density measurements, together with a rigorous understanding and quantification of their uncertainty, are essential.

Several methods are available for measuring the density of solid materials. For samples with a regular shape, the actual density can be calculated from the measured mass and the volume determined from the sample dimensions. This method is straightforward but is less suitable for irregularly shaped samples or with rough surfaces, where determining the volume from the dimensions can be difficult and introduce additional uncertainty. Furthermore, small samples are challenging to measure as the accuracy is limited by the uncertainties of the balance and caliper or gauge. Displacement methods measure the skeletal density by determining the volume of fluid displaced by the sample. Pycnometry is one example, in which the volume of a solid is determined using the displacement of a liquid or gas.[13,14] Gas pycnometry commonly uses helium because its small atomic size allows it to access fine surface features and open interconnected pores. However, in highly porous materials, helium can penetrate the pore network and access a portion of the internal pore volume. This reduces the measured volume of the solid and leads to an overestimation of the actual and skeletal density. Another approach is the Archimedes method, which determines the sample volume from the buoyant force acting on a sample immersed in a liquid.[13,15] This method is particularly useful for irregularly shaped solid samples because the volume does not need to be determined from the sample dimensions.

In an Archimedes measurement, the sample is weighed in air and then while immersed in a liquid; the difference between the two measurements is used to determine the volume of the sample. When the density of the liquid is known, the skeletal density can be calculated from the measured masses and the liquid density.[15] Hydrostatic weighing based on this principle has been used for high-precision density measurements of solid materials.[15,16] The Archimedes method, for example, is often applied for thermoelectric materials to measure the density or correct the electrical and thermal properties for the porosity of the sample.[17] However, small samples can have large uncertainty due to the relative uncertainty of mass measurements. Additionally, the liquid density can change with temperature and composition, making its accurate determination important for the overall measurement.[18,19,20] Ethanol is commonly used as the immersion liquid for Archimedes measurements, [17,21,22] but its density depends on factors such as water content and temperature, making direct measurement of the liquid density preferable to relying only on a tabulated value. Other experimental factors, such as trapped bubbles on the sample, changes in the composition of the immersion liquid over time, and variation in repeated mass measurements, can also affect the measured density.[15,22] These sources of uncertainty raise the question of how strongly each measurement contributes to the uncertainty in the calculated solid density and what measurement conditions are required to obtain a desired level of accuracy. At the same time, there is a need for density measurement systems that are not only accurate, but also accessible and readily reproducible between laboratories.

Three-dimensional (3D) printing has become an increasingly accessible approach for developing and producing laboratory equipment.[23,24] The ability to fabricate components directly from digital designs allows experimental systems to be readily customized, modified, and reproduced without the need for specialized manufacturing equipment.[25,26] This is particularly useful for measurement systems that may be implemented across different laboratories, as the same digital design can be shared and used to produce consistent components at each location.[23,27] In this way, 3D printing provides a practical means of improving the accessibility and reproducibility of experimental equipment while facilitating the adoption of standardization of measurement methods between laboratories.[26]

In this work, an easily accessible Archimedes-based system was developed using 3D printing for the accurate and precise measurement of both solid and liquid density. The system was evaluated through calibration and repeated measurements to assess its accuracy and precision. Given that the density of the liquid is directly used in the calculation of solid density, particular attention was given to its determination. An open-source graphical user interface (GUI) software was also developed to simplify the measurement process and automatically calculate the density and associated uncertainty from the experimental data. An uncertainty analysis was used to determine how the different measurement uncertainties contribute to the final density uncertainty. Furthermore, the relationship between sample mass and density uncertainty was investigated to establish the minimum sample mass required to achieve a desired density uncertainty.

# Methodology

**3D-Printed Apparatus**

A custom 3D-printed apparatus was designed to allow density measurements to be performed using an analytical balance (VWR-160TC) while minimizing the additional mass placed on the balance and maintaining compatibility with balances enclosed by a housing. Enclosures are vital to reduce the air flow and improve the accuracy of the measurements. The apparatus was designed as several separate components to allow for easy assembly and removal from the balance. A schematic of the assembled apparatus is shown in **Figure 1**. The apparatus consists of a bottom plate that fits directly onto the balance (8 cm diameter balance plate) and two vertical posts that fit into holes in the plate. A three-leg platform positioned around the weighing assembly of the balance supports the beaker and contains two additional holes through which the posts pass. The upper ends of the posts are connected by an attachment piece containing four holes, with two used to secure the posts and two used to hold copper wires (0.1 mm diameter wires). Copper wires were used as they are chemically resistant to most solutions, do not absorb or retain liquid, and maintain a stable length during measurements. The copper wires connect the bottom plate to the sample basket (1.5 mm diameter holes), allowing the sample to be suspended within the beaker during measurement. The present apparatus was fabricated using PET-G because of its ease of printing and straightforward fabrication. For applications involving harsher chemical solvents (e.g., acids or halogens), polyvinylidene fluoride (PVDF) is recommended due to its superior chemical resistance. The STL files for the 3D-printed components are available for download. An open-source GUI – Archimedes LS – was developed to distribute the Python code widely among the research community. No programming skills are required to determine the density with the associated uncertainties and can be freely downloaded from GitHub.

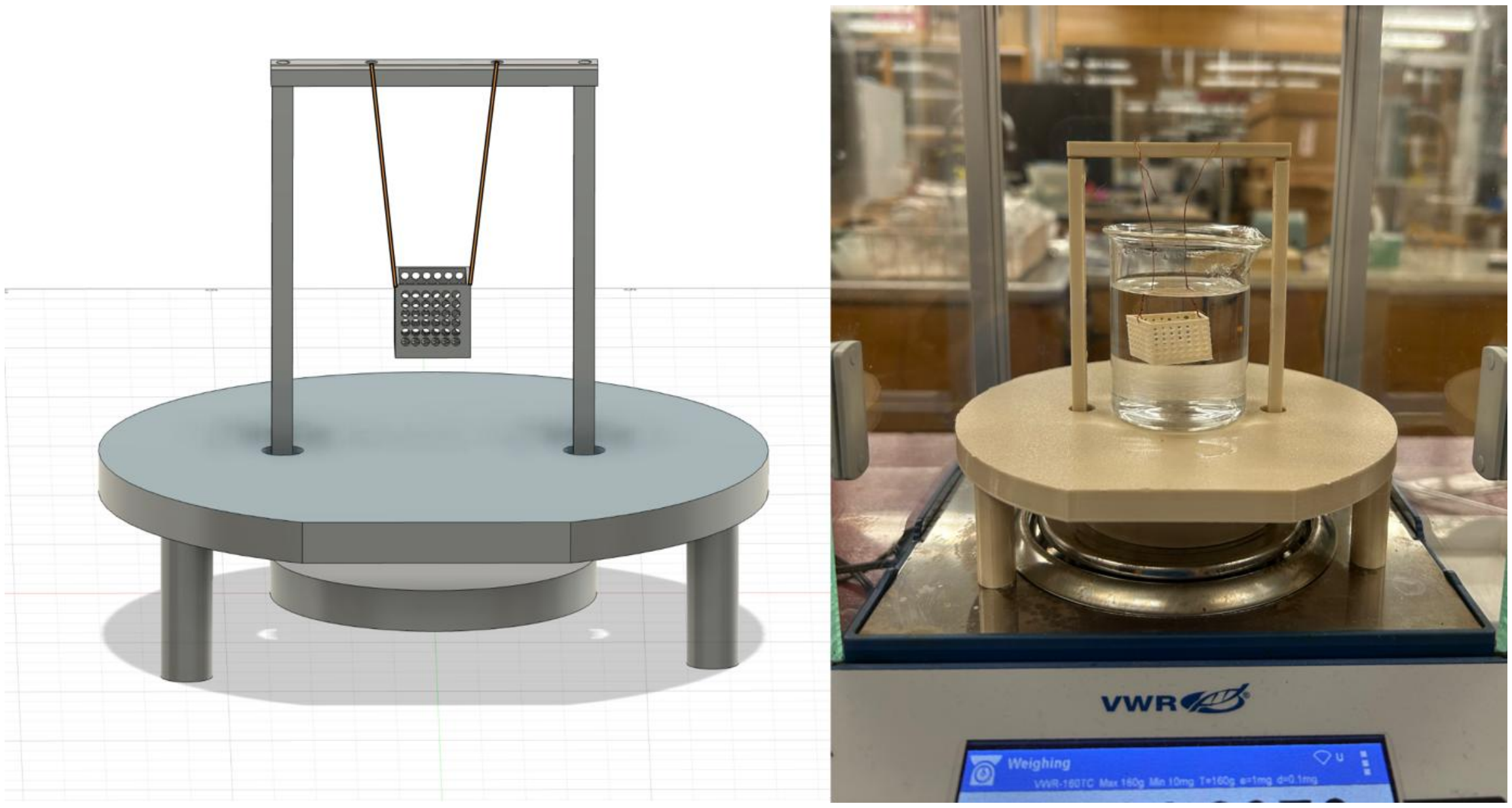

**Figure 1.** 3D-printed density measurement system apparatus.

**Calibration of Density Measurement System**

The liquid density was calibrated using copper (99.99%, BeanTown Chemicals), vanadium (99.95%, Sigma Aldrich), and indium (99.999+%, Sigma Aldrich) standards with known densities. Measurements were performed using 95% ethanol (Commercial Alcohols), hexanes (Fisher Chemical), and distilled water. For each measurement, the standard was weighed twice in air before the measurement system was placed on the balance. The standard was then placed in the sample basket and suspended in the liquid without contacting the liquid container. Temperature was monitored using an Xplorer GLS datalogger thermocouple (PASCO) during each liquid mass measurement and the recorded temperatures were averaged at the end of each measurement set. The known standard densities corresponding to the average measured temperatures were used in the density calculations. The apparatus was subsequently removed from the balance, and the procedure was repeated for a total of 10 measurements for each standard. The liquid was replaced after five measurements. The average masses measured in air and in liquid, along with their standard deviations, were calculated using Microsoft Excel and entered into the GUI analysis program to determine the density and the corresponding uncertainty of the liquid. The densities of the liquids measured using the Archimedes system were compared with values obtained using a 25 mL pycnometer (Kimble Glass Inc.) with distilled water as reference. The calibration procedure and results are discussed further in the Results and Discussion section.

**Density and Uncertainty Calculations**

The density of the calibration liquid, $\rho_l$, was determined from the slope of the densities of the calibration standards over the measured masses of the calibration standards according to

$$\rho_l = \rho_{standard} \frac{m_a - m_l}{m_a} \tag{1}$$

where $\rho_{standard}$ is the density of the calibration standard, $m_a$ is its average measured mass in air, and $m_l$ is its average apparent mass in the liquid. The uncertainty in the liquid density, $\sigma_{\rho_l}$, was determined by propagation of uncertainty using the standard deviations of the measured masses:

$$\sigma_{\rho_l} = \sqrt{\left(\frac{\delta \rho_l}{\delta m_a} \sigma_{m_a}\right)^2 + \left(\frac{\delta \rho_l}{\delta m_l} \sigma_{m_l}\right)^2} \tag{2}$$

with

$$\frac{\delta \rho_l}{\delta m_a} = \frac{\rho_{standard} m_l}{m_a^2} \tag{3}$$

and

$$\frac{\delta \rho_l}{\delta m_l} = \frac{-\rho_{standard}}{m_a} . \tag{4}$$

For comparison with the Archimedes measurements, the density of the liquid was also determined with a 25 mL gravity pycnometer. The specific gravity (SG) of the liquid was calculated according to

$$SG = \frac{M_w - M_0}{M_s - M_0} \tag{5}$$

where $M_s$ is the mass of the pycnometer containing the sample liquid, $M_w$ is the mass of the pycnometer containing distilled water, and $M_0$ is the mass of the empty pycnometer. The density of the sample liquid was then calculated using

$$\rho_l = SG \times \rho_{distilled\ water\ at\ test\ temperature} \tag{6}$$

where $\rho_{distilled\ water\ at\ test\ temperature}$ is the density of distilled water at the measured test temperature.

The calibrated liquid density was then used to determine the density of the solid according to

$$\rho_s = \frac{m_a \rho_l}{m_a - m_l} \qquad (7)$$

where $\rho_s$ is the unknown density of a solid. The uncertainty in the solid density, $\sigma_{\rho_s}$, was determined using propagation of uncertainty from the uncertainties in the air mass, liquid mass, and liquid density:

$$\sigma_{\rho_s} = \sqrt{\left(\frac{\delta \rho_s}{\delta m_a}\sigma_{m_a}\right)^2 + \left(\frac{\delta \rho_s}{\delta m_l}\sigma_{m_l}\right)^2 + \left(\frac{\delta \rho_s}{\delta \rho_l}\sigma_{\rho_l}\right)^2} \qquad (8)$$

with

$$\frac{\delta \rho_s}{\delta m_a} = -\frac{m_l \rho_l}{(m_a - m_l)^2}\,, \qquad (9)$$

$$\frac{\delta \rho_s}{\delta m_l} = \frac{m_a \rho_l}{(m_a - m_l)^2}\,, \qquad (10)$$

and

$$\frac{\delta \rho_s}{\delta \rho_l} = \frac{m_a}{m_a - m_l}\,. \qquad (11)$$

These partial derivatives were used to evaluate the contribution of each uncertainty to the final solid density uncertainty. The relative uncertainty in the measured solid density was defined as

$$relative\ uncertainty = \frac{\sigma_{\rho_s}}{\rho_s} \cdot 100\%\,. \qquad (12)$$

The minimum sample mass required to determine the density within a specific uncertainty can be calculated using

$$m_{min} = \rho_s \sqrt{\frac{(\rho_s - \rho_l)^2 \sigma_{m_a}^2 + \rho_s^2 \sigma_{m_l}^2}{\rho_l^2 \sigma_{\rho_s}^2 - \rho_s^2 \sigma_{\rho_l}^2}} \qquad (13)$$

where $m_{min}$ is the minimum sample mass required to achieve the specified relative uncertainty. The overall analytical and experimental workflow is summarized in **Figure 2**.

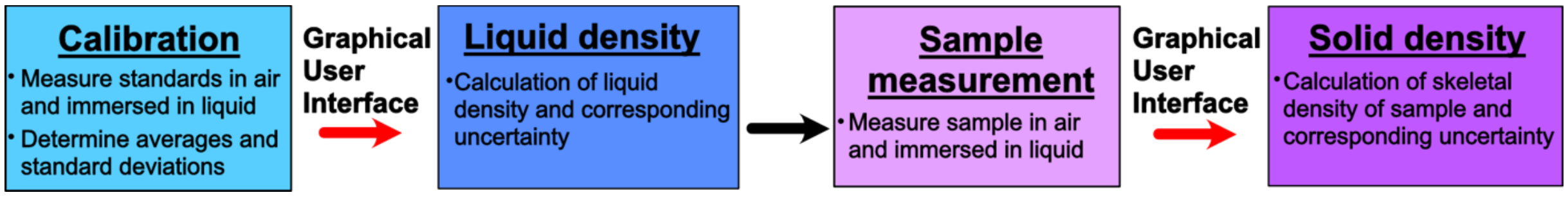

**Figure 2.** Overview of the experimental and analytical workflow used to determine liquid and skeletal densities and associated uncertainties.

# Results and Discussions

The liquid densities of ethanol, distilled water, and hexanes were calibrated using copper, vanadium, and indium standards with known densities; masses are given in **Tables 1-3**. The average temperature during the measurements was approximately 20.5 ± 2.3 °C; therefore, the known densities of the solid standards at 20 °C; 8.935 g/cm$^3$ for copper,[28] 6.099 g/cm$^3$ for vanadium,[28] and 7.29 g/cm$^3$ for indium,[28] were used in the calculations. The measured liquid densities were 0.8116 ± 0.0005 g/cm$^3$ for ethanol, 0.9982 ± 0.0013 g/cm$^3$ for distilled water, and 0.6748 ± 0.0016 g/cm$^3$ for hexanes, as shown in **Figure 3 (a)**. All three measured densities showed good agreement with the expected literature values; 0.8114 g/cm$^3$ for ethanol,[29] 0.9981 g/cm$^3$ for water,[29] 0.665 - 0.675 g/cm$^3$ for hexanes,[30] with the density of distilled water showing particularly close agreement. The measured density of hexanes was slightly higher than the expected value, which may be related to the commercial solvent containing a mixture of isomers and differences in composition between the solvent and the reference value. Generally, the composition of commercial solvents can also change after a bottle is opened due to factors such as evaporation, contamination, or absorption of moisture, volatile organic compounds (VOC), or other molecules from the surrounding environment. Opening a fresh bottle of solvent for every measurement would help minimize these changes; however, doing so would substantially increase the cost. As a result, the measured density may differ from the reported literature value. Temperature is also a factor that can affect the measured liquid density. The density of distilled water decreases from approximately 0.9981 g/cm$^3$ at 20 ºC to 0.9970 g/cm$^3$ at 25 ºC, corresponding to a change of 0.12% over a 5 ºC increase.[31] Although this change is relatively small, it demonstrates that density measurements are temperature dependent. Therefore, the temperature at which a density measurement is taken should be considered when comparing experimental results with literature values, particularly when the measurements are conducted at different temperatures.

The density of distilled water was subsequently used as a reference for comparison with measurements obtained using the 25 mL gravity pycnometer. The pycnometer measurements gave densities of 0.8102 g/cm$^3$ for ethanol and 0.6732 g/cm$^3$ for hexanes, corresponding to differences of 0.18% and 0.24% respectively, from the Archimedes measurements. The variation in the individual mass measurements was then examined for the calibration standards. Vanadium exhibited the largest variation in mass measurements, followed by indium and copper, consistent with the lower mass of the vanadium standard. The standard deviations were also generally higher when the standards were measured while immersed in the liquid compared with measurements made in air, indicating greater variability in the liquid measurements (**Table 1 – 3**).

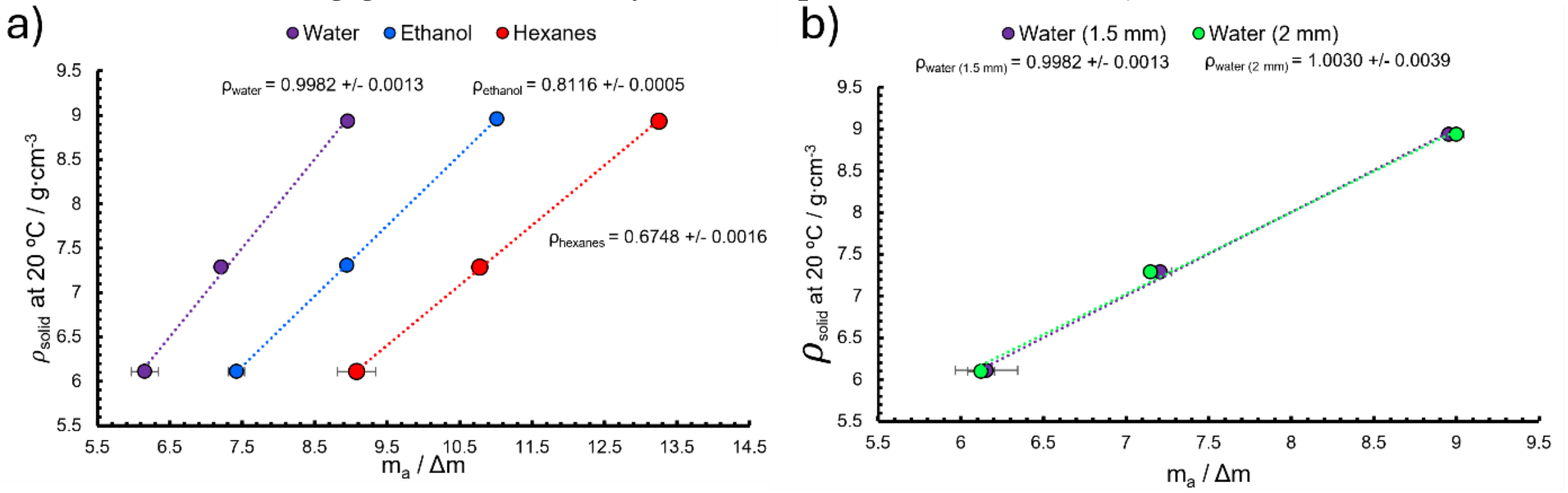

**Figure 3.** Calibration of the Archimedes density measurement system using a) ethanol, distilled water, and hexanes and b) distilled water using baskets with 1.5 and 2 mm diameter holes. Density of solid standards at 20 °C is plotted as a function of the change in mass between measurements in air and in liquid.

The effect of hole size of the basket on the measured liquid and solid densities was investigated by comparing baskets with 1.5 mm and 2 mm diameter holes using distilled water, as shown in **Figure 3 (b)**. The basket with 1.5 mm holes produced a measured water density of 0.9982 ± 0.0013 g/cm$^3$, while the basket with 2 mm holes produced a density of 1.0030 ± 0.0039 g/cm$^3$. The larger-holed basket therefore resulted in a slightly higher measured density and a greater uncertainty. Note, the liquid densities of both hole sizes are in uncertainty. The effect of basket hole size on the calibration standard was also examined. The mass measurement uncertainties for vanadium and indium were lower when using the 2 mm basket, whereas copper showed greater uncertainty with the 2 mm basket. The larger size of the copper standard relative to the basket opening may have contributed to this difference, as the standard only just fits within the basket. The larger holes may also allow trapped air bubbles to escape more readily during immersion, which could affect the measured immersed mass, and consequently, the calculated density. Overall, the results indicate that basket hole size can slightly influence the measured density and measurement uncertainty, although the effect differs between the calibration standards.

**Table 1.** Experimental data from distilled water measurements.

| Element | $m_a$ (g) | $\sigma_{m_a}$ (g) | $m_l$ (g) | $\sigma_{m_l}$ (g) | $\rho_l$ exp. (g/cm$^3$) | $\rho_l$ ref. (g/cm$^3$)[30] |
|---|---|---|---|---|---|---|
| Cu | 12.4603 | $1.63 \cdot 10^{-4}$ | 11.0687 | $1.80 \cdot 10^{-3}$ | | |
| In | 1.8906 | $3.75 \cdot 10^{-4}$ | 1.6283 | $2.60 \cdot 10^{-3}$ | 0.9982 | 0.9981 |
| V | 0.1008 | $9.44 \cdot 10^{-5}$ | 0.0844 | $5.40 \cdot 10^{-4}$ | | |

**Table 2.** Experimental data from ethanol measurements.

| Element | $m_a$ | $\sigma_{m_a}$ (g) | $m_l$ | $\sigma_{m_l}$ (g) | $\rho_l$ exp. (g/cm$^3$) | $\rho_l$ ref. (g/cm$^3$)[29] |
|---|---|---|---|---|---|---|
| Cu | 12.4604 | $1.45 \cdot 10^{-4}$ | 11.3290 | $7.65 \cdot 10^{-4}$ | | |
| In | 1.8916 | $2.21 \cdot 10^{-4}$ | 1.6801 | $5.02 \cdot 10^{-4}$ | 0.8116 | 0.8114 |
| V | 0.1008 | $5.87 \cdot 10^{-5}$ | 0.0872 | $1.63 \cdot 10^{-4}$ | | |

**Table 3.** Experimental data from hexanes measurements.

| Element | $m_a$ | $\sigma_{m_a}$ (g) | $m_l$ | $\sigma_{m_l}$ (g) | $\rho_l$ exp. (g/cm$^3$) | $\rho_l$ ref. (g/cm$^3$)[30] |
|---|---|---|---|---|---|---|
| Cu | 12.4604 | $2.66 \cdot 10^{-4}$ | 11.0759 | $7.11 \cdot 10^{-3}$ | | |
| In | 1.8899 | $5.59 \cdot 10^{-4}$ | 1.6256 | $1.56 \cdot 10^{-3}$ | 0.6748 | 0.665 - 0.675 |
| V | 0.1008 | $9.94 \cdot 10^{-5}$ | 0.0844 | $2.67 \cdot 10^{-4}$ | | |

A 0.9990 cm$^3$ tungsten cube (0.996 cm x 1.001 cm x 1.002 cm, 99.95% purity) weighing 19.1218 g was used to compare the density obtained from geometric measurements with that measured using the Archimedes system. The actual density was calculated using $\rho = m/V$, where $m$ is the mass and $V$ is the volume, giving an actual density of 19.1414 g/cm$^3$, while the Archimedes system measured a skeletal density of 19.2922 ± 0.0147 g/cm$^3$. The Archimedes measurement is closer to the reference density of tungsten; 19.3 g/cm$^3$ at 20 °C.[32] The lower density obtained from the geometric measurements may be attributed to the engraved features/surface pores of the cube. The geometric method uses external dimensions to calculate the volume and therefore does not account for surface pores or other voids, which can result in an underestimated actual density.

The effect of sample mass on the uncertainty of the measured solid density was investigated. The relative uncertainty in the calculated density was determined for different sample masses, and the resulting relationship is shown in **Figure 4**. Calculations were done with liquid densities of 0.6 g/cm$^3$ and 1.5 g/cm$^3$, representing approximate lower and upper densities of commonly used solvents. [33] As sample mass increased, the relative uncertainty in the measured density decreased. This indicates that larger samples reduce the relative contribution of the mass measurement uncertainty to the calculated density. The effect of balance uncertainty was also evaluated, with lower balance uncertainty resulting in lower uncertainty in the measured solid density. These results demonstrate the importance of both sample mass and balance precision in obtaining precise density measurements.

The analytical relationship between sample mass and density uncertainty was used to predict the minimum sample mass required to achieve a specified relative uncertainty. The effects of balance uncertainty, solid density, and liquid density on the calculated minimum sample mass were evaluated using **Equation 12**. A relative uncertainty of 1% was used as the target uncertainty and the resulting trends are shown in **Figure 4**. This target was selected based on the sensitivity of the predicted thermal conductivity and resistivity to porosity. The Klemens model was used to evaluate the effect of porosity on thermal conductivity according to $\kappa = \kappa_{exp} / (1\text{-}0.75p)$ with $\kappa_{exp}$ as the experimental (measured) thermal conductivity and $p$ as porosity, while the Maxwell model was used to evaluate the corresponding effect on resistivity according to $\rho = \rho_{exp}\cdot (1\text{-}p)/(1+0.5p)$ with $\rho_{exp}$ as the experimental (measured) electrical resistivity. To compare the effect of porosity on the two properties, the experimental thermal conductivity and resistivity were normalized to a value of 1 W $m^{-1}$ $K^{-1}$ or Ω·m. At a porosity of $p$ = 0.1, the Klemens model gives an intrinsic thermal conductivity of 1.0811 W $m^{-1}$ $K^{-1}$, while the Maxwell model gives an intrinsic resistivity of 0.8571 Ω·m. If the porosity varies by ± 0.03, corresponding to a range of $p$ = 0.07 – 0.13, the predicted thermal conductivity changes by approximately – 2.8% to +2.4 %, while the predicted resistivity changes by approximately -4.7% to +4.9%. In comparison, a smaller porosity range of $p$ = 0.09 – 0.11 (± 0.01) results in a thermal conductivity change of ~ ± 0.8% and resistivity of ~ ± 1.6%. This comparison shows that greater uncertainty in porosity can lead to noticeable changes in the predicted thermal and electrical properties. Therefore, reducing the uncertainty in the measured density is important for obtaining more reliable values and, consequently, more reliable property predictions.

The minimum sample masses calculated using the analytical expression (**Eq. 13**) were identical to those obtained using the analytical software, confirming agreement between the two approaches. **Figure 4 (a)** shows that the minimum sample mass decreases with the balance uncertainty from 5 to 2 and finally, 1 mg. Similarly, if the balance uncertainty remains the same the minimum sample mass is linearly proportional to the solid densities. Denser compounds require a higher sample mass or a lower mass uncertainty to achieve the same relative uncertainty as a lower dense sample because the Archimedes depends on the volume displacement by the liquid.

Increasing the liquid density from 0.6 g/$cm^3$ (**Figure 4 (a)**) to 1.5 g/$cm^3$ (**Figure 4 (b)**) also reduced the required sample mass. The results also show that the required balance precision depends on the density of the liquid. At higher liquid densities, a larger mass measurement uncertainty can be tolerated while still achieving the same target uncertainty in solid density (**Figure 4 (c)**). The effect is particularly beneficial for lower-density solids, for which using a higher-density liquid allows a lower overall uncertainty (**Figure 4 (d)**). The analytical expression further showed that the uncertainty in the immersed mass measurement has a greater influence on the minimum sample mass than the uncertainty in the air mass measurement. For a valid solution, the denominator of **Equation 13** must be greater than zero, requiring

$$\sigma_{\rho_s} > \frac{\rho_s}{\rho_l}\sigma_{\rho_l}. \tag{14}$$

This condition shows that higher liquid density and lower liquid density uncertainty are favorable for obtaining a valid and lower minimum sample mass. While often ethanol or isopropanol is used as liquid for the Archimedes method, their relatively low densities increase the uncertainty in the measured sample density. When the sample is compatible with water, its higher density can improve the precision of the measurements. Furthermore, higher-density organic solvents such as chlorobenzene or chloroform can improve the measurements. However, the chemical compatibility of the solvent must be evaluated not only for the sample material but also for the components immerged in the solvent such as the basket and wires. For example, the present apparatus consists of PET-G which is not compatible with chloroform and hence, the printing filament has to be replaced with chemically resistant material for such measurements.

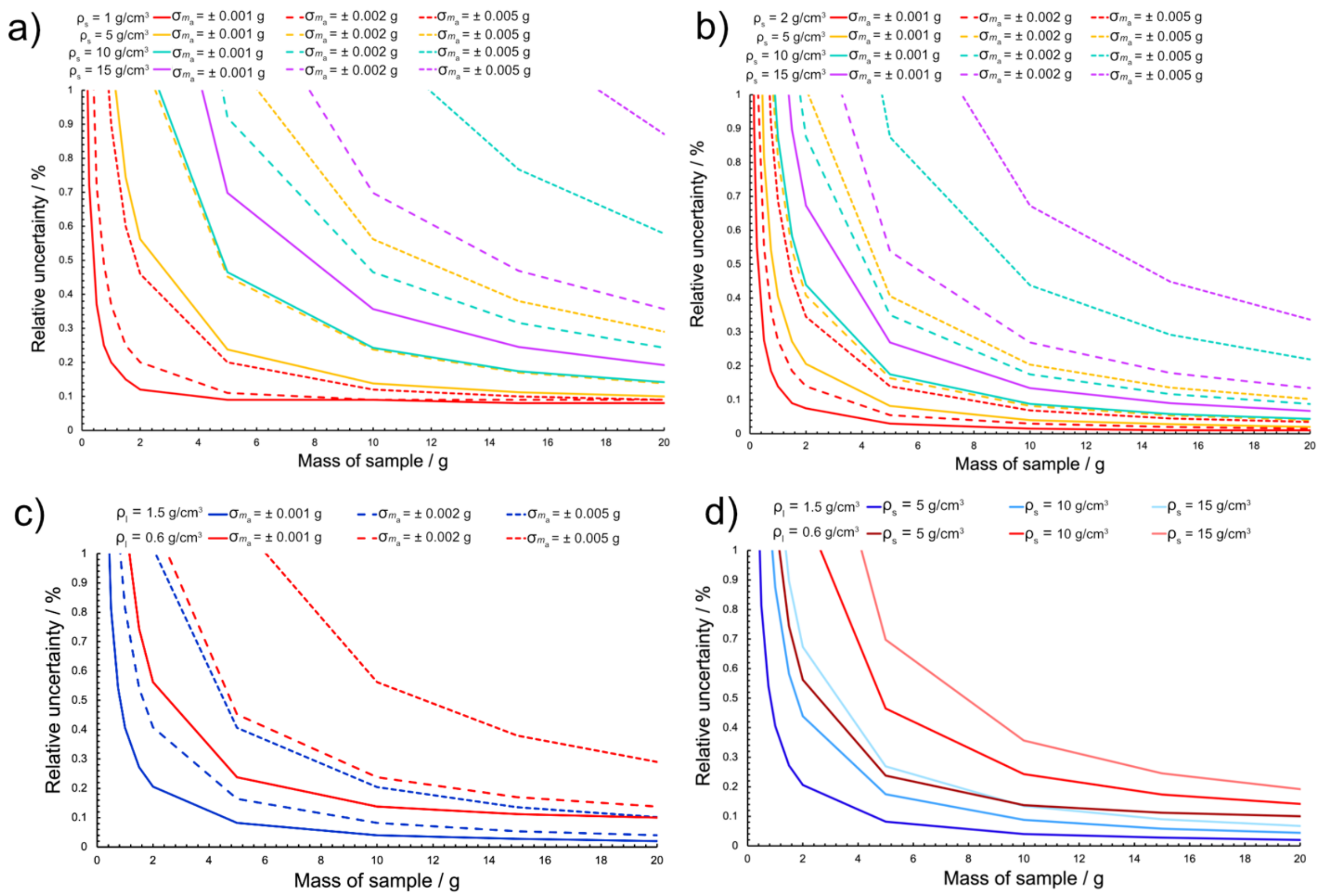

**Figure 4.** Calculated relative uncertainty in solid density as a function of sample mass to determine the minimum mass. Lighter sample masses increase the relative uncertainties. Densities of the material and its corresponding mass uncertainties are varied while the liquid density is held constant at a) 0.6 g/cm$^3$ and b) 1.5 g/cm$^3$. c) The density of the liquid and its associated uncertainties are changed while holding the solid density constant at 5 g/cm$^3$. d) Changing both solid and liquid densities with keeping the uncertainty (± 0.001 g) constant. Note, colors indicate a specific density whereas solid, dashed, and dotted lines represent a specific uncertainty.

# Conclusions

A low-cost, accessible Archimedes density measurement system was developed using 3D-printed components and the liquid densities (ethanol, distilled water, and hexanes) was successfully measured using copper, vanadium, and indium standards, showing good agreement with literature. The close agreement between the Archimedes and pycnometer measurements further supports the accuracy of the system. Accurate measurements of the liquid density are essential to calibrate the system for the evaluation of solid densities. The results show that larger sample masses, lower balance uncertainty, and higher density liquids reduce the relative uncertainty in the calculated solid density, while basket hole size can affect both the measured density and its uncertainty. In addition to the 3D-printed Archimedes apparatus, an open-source GUI (Archimedes LS) was introduced to quickly determine the density and the associated densities, including the determination of the minimum mass for a desired uncertainty. It was found that the analytical expression for minimum sample mass agreed with the software calculations, providing a useful method for selecting measurement conditions based on a desired uncertainty. As the apparatus and software are easily accessible for a wide range of researchers, the present method can become a standard for accurate and precise density measurements with an advanced understanding of the uncertainty, enhancing the comparability and consistency of density measurements among laboratories.

# Data availability

The source code is available at a public GitHub repository (https://github.com/JanPohls/ArchimedesLS) and the folder including the executable of the Archimedes LS GUI can be downloaded from the website (https://carrollpohlslab.ext.unb.ca/software/). The GUI was tested on Windows10 and Windows11. The STL files can be also downloaded from the website (https://carrollpohlslab.ext.unb.ca/software/).

# Author contributions

E.D.: Data acquisition, data analysis, design of the 3D model, figures, writing – original draft. J.-H.P.: conceptualization, supervision, writing – review & editing.

# Conflicts of interest

There are no conflicts to declare.

# Acknowledgements

This work received funding from the Discovery Grant from the Natural Sciences and Engineering Research Council of Canada and the Canada Foundation for Innovation (CFI) for John R. Evans

Leaders Fund (JELF). Furthermore, we would like to acknowledge the support from the New Brunswick Innovation Foundation (NBIF) and ResearchNB.